# Simulated sulfur K-edge X-ray absorption spectroscopy database of lithium thiophosphate solid electrolytes

Haoyue Guo,[1,†,*] Matthew R. Carbone,[2,†,*] Chuntian Cao,[2] Jianzhou Qu,[1] Yonghua Du,[3] Seong-Min Bak,[3] Conan Weiland,[4] Feng Wang,[5,**] Shinjae Yoo,[2] Nongnuch Artrith,[1,6,7,*] Alexander Urban,[1,6,8,*] Deyu Lu [9,*]

[1] *Department of Chemical Engineering, Columbia University, New York, New York 10027, USA*
[2] *Computational Science Initiative, Brookhaven National Laboratory, Upton, New York 11973, USA*
[3] *National Synchrotron Light Source II, Brookhaven National Laboratory, Upton, New York 11973, USA*
[4] *Material Measurement Laboratory, National Institute of Standards and Technology, Gaithersburg, Maryland 20899, USA*
[5] *Interdisciplinary Science Department, Brookhaven National Laboratory, Upton, New York 11973, USA*
[6] *Columbia Center for Computational Electrochemistry, Columbia University, New York, New York 10027, USA*
[7] *Materials Chemistry and Catalysis, Debye Institute for Nanomaterials Science, Utrecht University, 3584 CG Utrecht, The Netherlands*
[8] *Columbia Electrochemical Energy Center, Columbia University, New York, New York 10027, USA*
[9] *Center for Functional Nanomaterials, Brookhaven National Laboratory, Upton, New York 11973, USA*

† Equally contributing authors
* Corresponding authors: Haoyue Guo (haoyue1619@gmail.com), Matthew R. Carbone (mcarbone@bnl.gov), Nongnuch Artrith (n.artrith@uu.nl), Alexander Urban (au2229@columbia.edu), and Deyu Lu (dlu@bnl.gov)
** Present address: Applied Materials Division, Argonne National Laboratory, 9700 S. Cass Avenue, Lemont, IL 60439

X-ray absorption spectroscopy (XAS) is a premier technique for materials characterization, providing key information about the local chemical environment of the absorber atom. In this work, we develop a database of sulfur K-edge XAS spectra of crystalline and amorphous lithium thiophosphate materials based on the atomic structures reported in *Chem. Mater.*, 34, 6702 (2022). The XAS database is based on simulations using the excited electron and core-hole pseudopotential approach implemented in the Vienna Ab initio Simulation Package. Our database contains 2681 S K-edge XAS spectra for 66 crystalline and glassy structure models, making it the largest collection of first-principles computational XAS spectra for glass/ceramic lithium thiophosphates to date. This database can be used to correlate S spectral features with distinct S species based on their local coordination and short-range ordering in sulfide-based solid electrolytes. The data is openly distributed via the Materials Cloud, allowing researchers to access it for free and use it for further analysis, such as spectral fingerprinting, matching with experiments, and developing machine learning models.

## Background & Summary

The glass/ceramic lithium thiophosphates (*gc*-LPS) along the composition line $Li_2S–P_2S_5$ are considered promising electrolytes for solid-state batteries because of their superionic lithium conductivity at room temperature ($>10^{-3}$ Scm$^{-1}$), soft mechanical properties, and low grain boundary resistance.[1,2] Although *gc*-LPS lacks long-range atomic ordering, it exhibits characteristic short-ranged structural motifs that vary with the LPS composition and can affect the Li conductivity.

**Figure 1** illustrates how the local coordination of S atoms with Li and P atoms in the crystalline phases of LPS changes with increasing $Li_2S$ content $x$ in $(Li_2S)_x(P_2S_5)_{1-x}$: P$_2$–S and P–S–Li$_2$ in LiPS$_3$ (($Li_2S)_{0.5}(P_2S_5)_{0.5}$); P$_2$–S–Li, P–S–Li$_2$, P–S–Li$_3$ and P–S–Li$_4$ in Li$_7$P$_3$S$_{11}$ (($Li_2S)_{0.7}(P_2S_5)_{0.3}$); P–S–Li$_2$, P–S–Li$_3$ and P–S–Li$_4$ in Li$_3$PS$_4$ (($Li_2S)_{0.75}(P_2S_5)_{0.25}$); P–S–Li$_3$, S–Li$_6$ and S–Li$_7$ in Li$_7$PS$_6$ (($Li_2S)_{0.875}(P_2S_5)_{0.125}$). To understand the short-range ordering and its impact on Li conductivity in *gc*-LPS, several characterization tools have previously been employed, including X-ray diffraction (XRD),[3–12] Raman spectroscopy,[3,10,12–18] nuclear magnetic resonance (NMR),[10,12,15,19–21] X-ray photoelectron spectroscopy (XPS),[16,22–30] and X-ray absorption spectroscopy (XAS).[16,24,31–34]

As **Figure 1** indicates, the average Li coordination number of S increases with the fraction of Li$_2$S in LPS, leading to a greater density of Li around S atoms. Such an increased local Li concentration has previously been argued to increase electron density and thereby give rise to a shielding effect around S atoms and a red shift of the S K-edge.[24] Tender energy XAS spectroscopy is therefore a natural choice to probe the local geometric and electronic structures in *gc*-LPS.

Sulfur is known to participate to a greater extent than phosphorus in interfacial reactions during cycling, forming $Li_2S$ at the negative electrode or other metal sulfides at the positive electrode (*e.g.*, NiS).[33] In contrast, phosphorus is mostly bound in the center of $PS_4^{3-}$ tetrahedra (as $P^{5+}$ species), except for the direct P–P bonding in $P_2S_6^{4-}$ motifs (as $P^{4+}$ species). There is no direct Li–P bonding in *gc*-LPS and hence sulfur is more sensitive to the change in Li stoichiometry. Based on these considerations, sulfur K-edge spectroscopy can be expected to yield important insights into the electrochemical reactions in LPS-based solid-state batteries.

Commonly, XAS spectra are interpreted by comparison with characteristic features in spectra taken from reference materials, however, this approach is challenging when the composition and structure of the material cannot be readily identified.[35] In order to aid with the interpretation of XAS measurements and to understand the nature of the short-range ordering and its impact on properties such as Li conductivity and the electronic structure, first-principles XAS simulations have previously been conducted.[35–43] These simulations involve the modelling of the excitation of a core electron into the conduction bands, leaving behind a core hole. Within methods based on density functional theory (DFT) band structure, two approaches are commonly used to account for the core-hole final state effect: *(i)* the excited electron and core-hole (XCH) method with self-consistent relaxation of valence electrons[44–46], which is implemented in, e.g., XSPECTRA[47] and the Vienna *Ab Initio* Simulation Package (VASP)[48], and *(ii)* many-body perturbation theory based on the Bethe-Salpeter equation treating the screening of valence electrons with linear response, which is implemented in, e.g., OCEAN[49] and EXCITING[50]. Generally, the many-body perturbation theory-based method is computationally more demanding.[35] In comparison, the XCH approach can provide a reasonable accuracy that is sufficient to compare trends with experimental measurements at a moderate computing time[36,44,51–54] and is therefore a good choice for the compilation of a large XAS database. In addition, Pascal *et al*. demonstrated that the XCH approach can reliably predict the features of the S K-edge of distinct coordination environments in Li-S batteries.[36]

XAS simulations have so far been limited to crystalline LPS phases, and to our knowledge no XAS simulations of *gc*-LPS have been reported owing to the complexity of the glassy phases. We recently mapped the phase diagram of *gc*-LPS by combining DFT, artificial neural network (ANN) potentials, genetic-algorithm (GA) sampling, and *ab initio* molecular dynamics (AIMD) simulations, to compile a database of stable and metastable *gc*-LPS atomic structures.[55] This *gc*-LPS phase diagram is the foundation for the herein reported database of simulated *gc*-LPS S K-edge XAS spectra.

Here, we report the S K-edge XAS simulations for an extensive database of LPS/*gc*-LPS structures. A workflow for automated calculations using the XCH approach (**Figure 2**) was implemented using the open-source Pymatgen package[56] and VASP.[48] The final database contains 2681 simulated S K-edge XAS spectra for 66 crystalline and glassy structures. Where possible, the simulated spectra were benchmarked by comparison with tender energy XAS spectroscopy measurements. The database is distributed via the Materials Cloud repository,[57] enabling open access by other researchers for further exploration. The workflow is available as **Supporting Information**, providing a tool for researchers to construct their own XAS spectral databases.

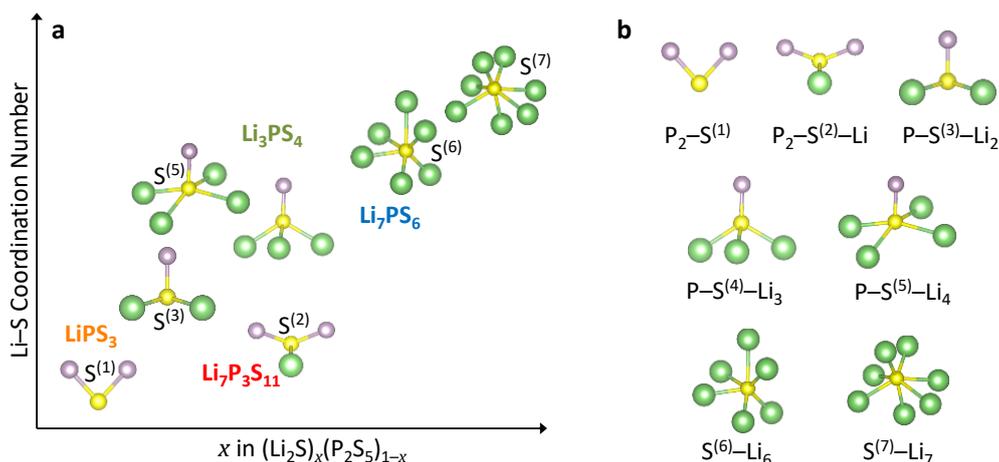

**Figure 1.** Schematic illustration of the local coordination of S atoms with Li and P atoms in selected $(Li_2S)_x(P_2S_5)_{1-x}$ crystalline structures. Li: green; S: yellow; P: purple. $LiPS_3$: orange region; $Li_7P_3S_{11}$: red region; $Li_3PS_4$: green region; $Li_7PS_6$: blue region.

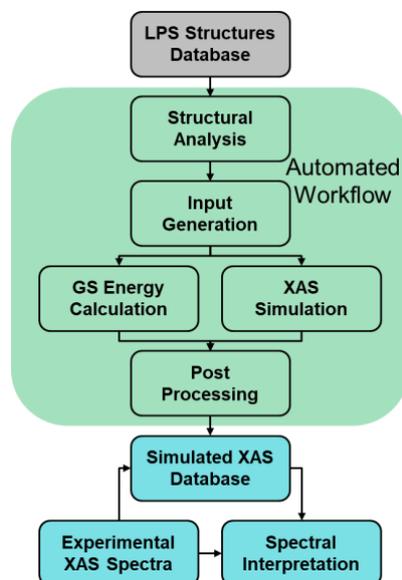

**Figure 2.** Flowchart with the workflow for building the sulfur K-edge XAS spectral database of crystalline and amorphous lithium thiophosphate materials.

## Methods

*Density Functional Theory Calculations*

All DFT calculations were carried out within the projector-augmented-wave (PAW) approach[58,59] as implemented in VASP.[58,60] The simulation parameters were carefully tested to ensure numerical convergence with the energy cut-off for the plane-wave basis set, the supercell size for XCH calculations, the density of the k-point meshes, and the number of unoccupied bands. Different exchange-correlation functionals, pseudopotentials from the VASP library, and core-hole charges (full vs. half) were compared to determine the optimal VASP input parameters for XAS simulations. The following parameters yielded converged results that compared best with experimental reference spectra.

Ground-state energy calculations and XAS simulations were performed with the local-density approximations (LDA) exchange-correlation functional[61] and the VASP GW pseudopotentials, which achieve a more accurate description of the post-edge region than regular LDA potentials because the GW pseudopotentials were optimized to yield more accurate scattering properties at high energies well above the Fermi level. A kinetic energy cut-off for the wavefunction of 400 eV, supercells with an edge length of at least 10~15 Å in each direction, and a full core hole were used for all calculations. Three times as many unoccupied bands as occupied bands were included in all calculations to ensure the convergence of the conduction band in the relevant energy range. Gaussian smearing with a width of 0.05 eV was used, and total energies were converged to better than $10^{-5}$ eV/atom. The first Brillouin zone was sampled using Γ-centered k-point meshes with the resolution of 0.25 $\text{Å}^{-1}$ generated with VASP's automatic sampling method. In XCH simulations, a constant Lorentzian broadening of 0.05 eV was introduced. Additional broadening can be added during post-processing when compared to experiment as discussed below.

In XAS simulations with the XCH approach, the final state is treated self-consistently subject to the presence of a core-hole in the S 1s orbital.[44,53] The XAS spectrum is calculated as the imaginary part of the frequency-dependent dielectric matrix averaged over the diagonal matrix elements within the PAW frozen-core approximation. For comparison with measured reference spectra, the simulated XAS spectra were convoluted with a Gaussian function with a full width at half maximum of 0.5 eV to simulate instrument broadening and with a Lorentzian function with an energy-dependent width of $0.59 \text{ eV} + a \times (E_c - E_{\text{cbm}})$ to simulate the core-hole lifetime broadening and quasiparticle life time broadening, where $a$ is a fitting parameter and $E_c$ and $E_{\text{cbm}}$ are DFT energy levels of conduction bands and conduction band minimum, respectively. Each absorption edge was aligned using the excitation onset determined from the total energy difference between the final state and the initial state, following a previously reported procedure.[54]

*Structure selection*
The structures for the XAS simulations were selected from the LPS structure library by Guo *et al.*[55] The *gc*-LPS structures in this dataset were generated by iterative manipulation of the known crystal structures along the $(Li_2S)_x(P_2S_5)_{1-x}$ composition line ($LiPS_3$, $Li_4P_2S_7$, $Li_7P_3S_{11}$, α-$Li_3PS_4$, β-$Li_3PS_4$, γ-$Li_3PS_4$, and $Li_7PS_6$) using a previously established protocol.[62,63] In short, (A) a supercell of a crystal structure was created, (B) either Li and S atoms were removed with a ratio of 2:1 ($Li_2S$), or P and S atoms were removed with a ratio of 2:5 ($P_2S_5$), and (C) low-energy configurations of the new composition were determined with a genetic (evolutionary) algorithm using an artificial neural network (ANN) interatomic potential as implemented in the atomic energy network (ænet) package.[64–66] For further details we refer the reader to reference.[55] From this dataset, the $(Li_2S)_x(P_2S_5)_{1-x}$ structures with the lowest formation energies relative to $Li_2S$ and $P_2S_5$ at each composition were chosen for XAS simulations. In addition, the above crystalline LPS compounds and the crystal structures of the sulfur-deficient $Li_2PS_3$ and $Li_{48}P_{16}S_{61}$ were included.

*Automated DFT workflow for constructing XAS database*
On the basis of the determined parameters from the benchmark systems, a workflow was devised for automated XCH calculations for generating an XAS database (**Figure 2**). For each optimized LPS structure, the workflow based on symmetry automatically determines the inequivalent S sites and their respective weights. Our implementation makes use of the symmetry tools from the Pymatgen package.[56] Pymatgen functions were further used to create supercells and generate VASP input files for single-point LDA calculations to obtain the ground state energy, and for XCH calculations for all symmetrically distinct S atoms in the supercell. Raw data from completed DFT calculations are post-processed, which mainly involves two steps: 1) applying the peak alignment to distinct S atoms using the excitation onset determined from the total energy difference between the final state and the initial state and 2) averaging the aligned spectra with the correct weights to compute the XAS spectrum of the whole system. Note that the data without averaging contains information about the XAS features of local atomic structures, which could be used for further exploration, *e.g.*, machine-learning-assisted spectral interpretation.

*Sample preparation and XAS measurements*
The experiment S K-edge XAS spectra were measured at the 8-BM and 7-ID-2 beamlines at National Synchrotron Light Source II (NSLS-II). The $P_2S_5$ spectrum was measured at 8-BM in fluorescence yield (FY) mode, and $Li_2S$, NiS, and β-LPS were measured at 7-ID-2 in electron yield (EY) mode. We used unfocused beam with spot sizes of 2.5 mm × 5 µm and 1 mm × 1 mm at 8-BM and 7-ID-2, respectively. Prior to the measurements, the samples were pressed into pellets with 1 cm diameter. For the 8-BM measurement, the sample was sealed between Kapton tape and polypropylene film in an argon-filled glovebox, and then transferred into the helium chamber at the beamline. For the 7-ID-2 measurements, the samples were mounted on a sample bar and sealed in an aluminized polymer bag in the glovebox, and then transferred into the vacuum chamber at the beamline using an argon-filled transfer bag. The experiment XAS spectra were processed with Athena software package.[67]

## Data Records

The database contains 66 structures with between 12 and 162 atoms, 18 of which are crystalline and the rest are amorphous (see Table 1). For each structure, a ground state self-consistent field (SCF) calculation is computed and stored in a directory named `input_SCF`. For every symmetrically inequivalent S site (between 1 and 86 per structure), a core-hole calculation is performed using the S core-hole pseudopotential. For each individual VASP calculation, we provide all input files except the pseudopotentials (`POTCAR` files) since those are distributed with VASP: `INCAR`, `POSCAR` and `KPOINTS`. That way, calculations can be rerun after reconstructing the appropriate potential file for each calculation. Due to the large size of many VASP output files, we only keep those necessary for presenting and reproducing the spectral database. These include the `INCAR`, `POSCAR` and `KPOINTS` input files, and the `OSZICAR` output file (to demonstrate the convergence of the calculation). Additionally, we save the Fermi energy where relevant (`efermi.txt`) and post-process the XAS from the `OUTCAR` (`mu.dat`; note: most regions with zero intensity are discarded to save space). Each spectrum consists of four columns: the energy, and the three components of the XAS (corresponding to the three polarization directions along the Cartesian coordinates). VASP 6.2.1 was used with GPU acceleration, and no post-processing was performed, such that the database is essentially preserved exactly as output by VASP. We provide short post-processing scripts for extracting key observables, such as the energy and spectral intensity. The spectral data are stored in the Materials Cloud (https://www.materialscloud.org).

**Table 1.** The construction of S K-edge XAS database in *gc*-LPS. Note that two compositions ($Li_3PS_4$ and $Li_4P_2S_7$) appear in both crystalline and glassy structure models, so the total number of compositions is 28 instead of 30.

| Database | Compositions | Structures | S Sites |
|---|---|---|---|
| Crystalline | 9 | 18 | 141 |
| Glassy | 21 | 48 | 2540 |
| Total | 28 | 66 | 2681 |

## Technical Validation

*Benchmark of the XAS simulations*
Our calculations started with the benchmark of the XAS simulations using reference sulfur compounds. Some of the most relevant compounds from *gc*-LPS/electrode interfacial degradation, including $Li_2S$, $P_2S_5$ and NiS, were selected as validation systems. To validate the VASP XAS simulations, the simulated spectra were compared against experimental measurements for three benchmark systems ($Li_2S$, $P_2S_5$ and NiS) as shown in **Figure 3**. The simulated spectra successfully reproduce the main features in reference systems. It is known that Kohn-Sham DFT underestimates band gaps and concomitantly band widths, due to inaccurate estimates of quasiparticle (excitation) energies based solely on the Kohn-Sham eigenspectrum.[36] Therefore, the calculated XAS spectra may underestimate peak separations compared to experiments, as seen in the Li$_2$S spectrum.

The XCH simulations successfully reproduce the three main features in the Li$_2$S spectrum at 2474, 2476.5, and 2484 eV, as well as the peak shoulder at 2480 eV, while the energy separation of the first two peaks was slightly underestimated. The relative intensities of these peaks are mostly successfully reproduced, with a slight underestimate of the intensity of the third peak at 2484 eV.

The spectrum of P$_2$S$_5$ exhibits a pronounced pre-edge feature. The structure of (P$_2$S$_5$)$_2$ is composed of two types of S atoms: 4 terminal S and 6 bridging S, denoted as S(1) and S(2), respectively in **Figure 3b** and indicated in structural motifs in **Figure 1**. The terminal S atom is coordinated with one P atom, and the P-S bond length is around 1.9 Å. In comparison, the bridging S atom is coordinated with two P atoms; the charge distribution over the bridging S is less negative than the terminal S, leading to a longer P-S bond length of 2.1 Å and blueshift of the absorption onset as shown in Figure 3b. Our results are the first demonstration that the pre-edge and main edge of the S K-edge in P$_2$S$_5$ can be attributed to two types of differently coordinated S atoms. This also demonstrates that XCH simulations can distinguish the inequivalent absorption sites.

To study the interfacial reaction between LPS and Ni-based cathodes, we also computed S K-edge XAS for NiS, a common degradation product of LPS in contact with Ni-based cathode materials, without and with a Hubbard $U$ correction of 3.9 eV to account for the correlation of the Ni $d$-band electrons. As shown in **Figure 3c**, the Hubbard $U$ correction does not change the main absorption edge, but leads to a broadened absorption edge and increased intensity in the post-edge region. While the $U$ value is dependent on the species and materials and must be tested carefully, the overall excellent agreement between the XCH simulations and experiments on exemplary reference compounds demonstrates the robustness of our approach and the reliability of our dataset.[41]

*Validation of DFT calculated S K-edge in crystalline β-Li$_3$PS$_4$.*
To further validate the simulated XAS spectra of the sampled *gc*-LPS phases, we also conducted experimental measurements of XAS reference data for LPS crystal structures. The S K-edge in crystalline *β*-Li$_3$PS$_4$ was measured under fluorescence mode at NSLS-II. As shown in **Figure 3d**, our computed XAS spectra are in excellent agreement with the experimental data. The absorption edge is around 2471 eV, which is likely due to the S 1s to S 3p $\sigma^*$ transition (dumbbell-shaped S$_2^{2-}$).[24] The simulated XAS not only reproduces most features, but also yields a comparable peak splitting for the S K-edge in the *β*-Li$_3$PS$_4$ crystal. In *β*-Li$_3$PS$_4$, there are three inequivalent S sites (denoted as S$^{(3)}$, S$^{(4)}$, S$^{(5)}$ in **Figure 1 and 3d**), where the local coordination is shown in **Figure 1**. While the charge distribution at the three S sites is comparable, the P–S and Li–S bond lengths exhibit a sizable variation. In this case, the core-level chemical shift cannot be simply explained by the bond length and charge transfer. It is an interesting future research direction to develop optimal structural and chemical descriptors for the interpretation of XAS spectral features in *gc*-LPS and especially the interphases at solid-state interfaces with Li metal anodes and Ni-based cathodes.

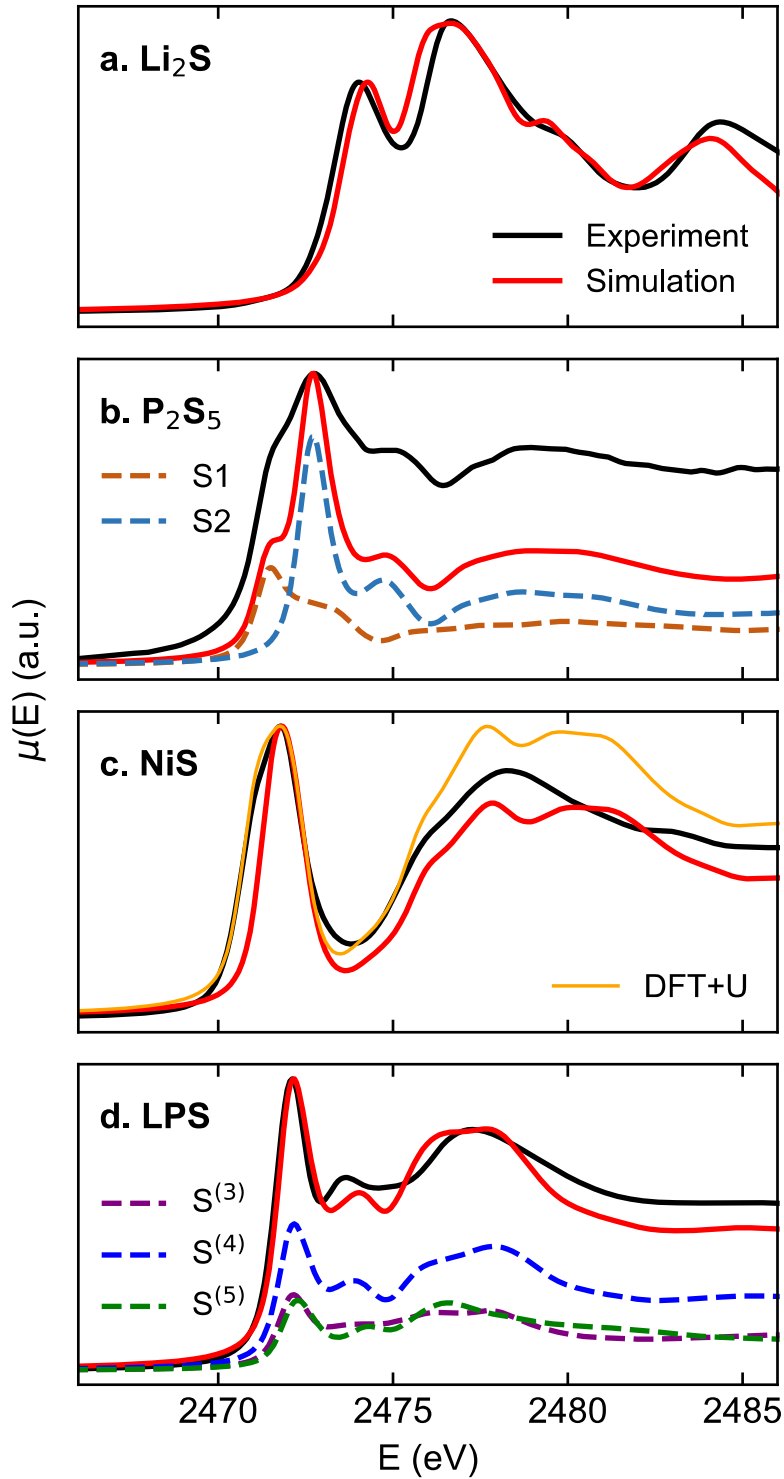

**Figure 3.** Benchmark results of XAS simulations of (a) Li$_2$S, (b) P$_2$S$_5$, (c) NiS, and (d) β-Li$_3$PS$_4$. In each subfigure, the black curve indicates the experimental spectrum, and the red curve indicates the simulated spectrum. Spectra calculated for different distinct S sites are shown as dashed lines in (b) and (d). The experimental spectra in (a, c, d) were measured in electron yield mode at Beamline 7-ID-2 of NSLS-II. The experimental spectrum in (b) was measured in fluorescence yield (FY) mode at Beamline 8-BM of NSLS-II, where the peaks of the P$_2$S$_5$ experimental spectrum are damped due to self-absorption in FY mode. DFT calculations with (orange) and without Hubbard $U$ correction for Ni (red) are shown in (c). The labels of different S sites in (d) correspond to the local motifs in Figure 1b.

## Usage Notes
See Code Availability.

## Code Availability
Short scripts used for extracting useful information from the VASP output files, such as the XAS and energies, are provided with the database.

## Data Availability
The data set can be obtained from the Materials Cloud ([doi.org/10.24435/materialscloud:92-0a](doi.org/10.24435/materialscloud:92-0a)) and contains VASP input and output files, Python scripts, metadata, and processed data.

## Acknowledgements
We acknowledge financial support by the U.S. Department of Energy (DOE) Office of Energy Efficiency and Renewable Energy, Vehicle Technologies Office, Contract No. DE-SC0012704. The research used the theory and computational resources of the Center for Functional Nanomaterials and Beamlines 7-ID-2 and 8-BM of NSLS-II, which are the U.S. DOE Office of Science User Facilities, and the Scientific Data and Computing Center, a component of the Computational Science Initiative, at Brookhaven National Laboratory under the Contract No. DE-SC0012704. We also acknowledge computing resources from Columbia University's Shared Research Computing Facility project, which is supported by NIH Research Facility Improvement Grant 1G20RR030893-01, and associated funds from the New York State Empire State Development, Division of Science Technology and Innovation (NYSTAR) Contract C090171, both awarded April 15, 2010.

Disclaimer: Commercial equipment, instruments, or materials are identified in this paper to specify the experimental procedure adequately. Such identification is not intended to imply recommendation or endorsement by the National Institute of Standards and Technology, nor is it intended to imply that the materials or equipment identified are necessarily the best available for the purpose.

## Author contributions
Each author's contribution to the work should be described briefly, on a separate line, in the Author Contributions section.

H.G.: structure selection, DFT calculations, benchmarking and parameter optimization of the EXC simulations, workflow conception, writing – initial draft and editing.

M.R.C.: workflow implementation, DFT and spectral calculations, analysis, writing – review and editing.

C.C., Y.D., S.B., C.W.: experimental XAS data acquisition.

J.Q.: workflow implementation.

F.W.: project conception, supervision.

N.A.: project conception, structure selection, workflow conception and implementation, writing – review and editing.

A.U.: conception, implementation of the workflow, writing – review and editing.

D.L.: project conception, supervision, writing – review and editing.

## Competing interests
The authors declare no competing interests.